\documentclass[a4paper,12pt]{article}
\usepackage{fullpage}

\begin{document}

\author{V.Yu.Irkhin$^{*}$ and M.I.Katsnelson \\
Institute of Metal Physics, 620219 Ekaterinburg, Russia}
\title{Electron spectrum, thermodynamics and transport in antiferromagnetic metals
at low temperatures}
\date{}
\maketitle

\begin{abstract}
Electron spectrum of $2D$ and $3D$ antiferromagnetic metals is calculated
with account of spin-fluctuation corrections within perturbation theory in
the $s-f$ exchange model. Effects of the interaction of conduction electrons
with spin waves in thermodynamic and transport properties are investigated.
At lowest temperatures $T<T^{*}\sim (\Delta /E_F)T_N$ ($\Delta $ is the AFM
splitting of the electron spectrum) a Fermi-liquid behavior takes place, and
non-analytic $T^3$ln$T$-contributions to specific heat are present for $D=3$%
. At the same time, for $T>T^{*}$, in $2D$ and ``nested'' $3D$ systems the
picture corresponds to a marginal Fermi liquid ($T$ln$T$-contributions to
specific heat and nearly $T$-linear dependence of resistivity). Frustrations
in the spin system in the $3D$ case are demonstrated to lead to similar
results. The Kondo contributions to electronic properties are analyzed and
demonstrated to be strongly suppressed. The incoherent contributions to
transport properties in the presence of impurity scattering are considered.
In particular, in the $2D$ case $T$-linear terms in resistivity are present
up to $T=0,$ and thermoelectric power demonstrates the anomalous $T\ln T$%
-dependence.
\end{abstract}

\section{Introduction}

The theory of electronic structure of highly correlated systems is up to now
extensively developed. The interest in this problem has grown in connection
with studying anomalous rare-earth and actinide compounds (e.g.,
heavy-fermion systems) and high-$T_c\,$superconductors (HTSC). Last time, a
possible formation of states, which differ from the usual Fermi liquid, is
extensively discussed. The non-Fermi-liquid (NFL) behavior of the excitation
spectrum up to lowest energies is now reliably established in the
one-dimensional case (the ``Luttinger liquid'' \cite{Mah}). However,
Anderson \cite{And} assumed occurrence of a similar situation in some
two-dimensional ($2D$) and even three-dimensional ($3D$)$\,\,$systems with
strong electron correlations, various mechanisms (resonating valence bond
(RVB) state, scattering anomalies, Hubbard's splitting etc.) being proposed.
Recently, an attempt has been made \cite{far} to revise the general
formulation of the Luttinger theorem (conservation of the volume under the
Fermi surface for arbitrary law of vanishing of electron damping at $E_F$)
which is a basis of the Fermi liquid description.

Another approach to the problem of NFL behavior was proposed by Varma et al
\cite{Var}. To describe unusual properties of HTSC (e.g., the $T$-linear
dependence of resistivity), these authors put forward a phenomenological
``marginal Fermi liquid'' (MFL) theory where electron damping is linear in
energy $E\,$(referred to the Fermi level), and the effective mass is
logarithmically divergent at $E\rightarrow 0$. Such a behavior was supposed
to result from the interaction with local Bose excitations which possess a
peculiar (linear in their energy and weakly $q$-dependent) spectral density.
Further the MFL theory was developed in a number of papers (see, e.g., \cite
{Zim,Wang,Sire}). In a simplest way, the MFL electron spectrum can be
reproduced in some crossover energy region for interacting electron systems
under the requirement of almost perfect nesting in $2D$ case \cite
{Vir,Gal,Zl,Jac}, which seems to be too strict for real systems. Similar
results were obtained with account of antiferromagnetic (AFM) spin
fluctuations in the vicinity of the AFM instability \cite{Kampf}. Almost $T$%
-linear behavior of the resistivity in paramagnetic $2D\,\,$metals with
strong AFM fluctuations was obtained by Moriya et al \cite{Mor} in a broad
temperature region.

Besides HTSC, a NFL behavior in some temperature intervals was found
experimentally in a number of uranium and cerium systems (U$_x$Y$_{1-x}$Pd$%
_3 $ \cite{UYPd}, UPt$_{3-x}$Pd$_x$ \cite{UPtPd}, UCu$_{5-x}$Pd$_x$ \cite
{UCuPd}, CeCu$_{6-x}$Au$_x$ \cite{CeCu}, U$_x$Th$_{1-x}$Be$_{13}$, Th$_{1-x}$%
U$_x$Ru$_2$Si$_2$, Ce$_x$La$_{1-x}$Cu$_2$Si$_2$ \cite{Map,Schub}). In
particular, $T\ln T\,$-term in electronic specific heat, $T$-linear
corrections to resistivity (both positive and negative ones), unusual
power-law or logarithmic $T$-dependences of magnetic susceptibility etc.
were observed \cite{Map}. This behavior is as a rule interpreted within the
two-channel Kondo scattering mechanism \cite{Tsv}, Griffith's point
mechanism \cite{Castr}, etc. At the same time, in a number of systems (UCu$%
_{5-x}$Pd$_x$, CeCu$_{6-x}$Au$_x$, U$_x$Y$_{1-x}$Pd$_3$) the NFL behavior
correlates apparently with the onset of antiferromagnetic (AFM) ordering
\cite{CeCu,Wu}.

An interesting behavior demonstrates the system is Y$_{1-x}$Sc$_x$Mn$_2$
\cite{Wada,Bal}. YMn$_2$ is an itinerant AFM with a frustrated magnetic
structure, and for $x=0.03$ (or under pressure) the long-range magnetic
order is suppressed and the $\,$linear specific heat is giant ($\gamma =140$
mJ/mol K$^2$). This system demonstrates strong anomalies of transport
properties (in particular, deviations from the quadratic $T$-dependence of
resistivity) \cite{YMNR}. A hypothesis about formation of a spin-liquid
state in this system was put forward in \cite{IKPL}. A detailed analysis of
the neutron scattering in this compound seems to confirm this hypothesis
\cite{lac1}.

Most of systems under consideration possess peculiarities of band and
spin-fluctuation specta. The Fermi surfaces of the anomalous $f$-systems
have complicated forms with several pieces \cite{zw}; we have to take into
account that for some of these pieces the ``nesting'' condition can hold. It
will be shown below that such cases need separate consideration (roughly
speaking, in the nesting situation the effective dimensionality of the
system diminishes by unity). Despite $3D$ picture of electron spectrum, spin
fluctuations in such systems as CeCu$_{6-x}$Au$_x$, CeRu$_2$Si$_2$
demonstrate $2D$-like behavior \cite{Lohn,Stock}.

Since practically all the above-discussed highly-correlated systems are
characterized by pronounced local moments and spin fluctuations, a detailed
treatment of the electron-magnon mechanism seems to be important to describe
their anomalous properties. In the present work we consider effects of
interaction of conduction electrons with usual spin-wave excitations in
metallic antiferromagnets with localized magnetic moments within the $s-d(f)$
exchange model. The latter condition results in that spin waves are well
defined in a large region of the $\mathbf{q}$-space. From the general point
of view, presence of the localized-electron system is favorable for the
violation of the Fermi-liquid picture. Indeed, we shall demonstrate that a
number of physical properties of the AFM metals with $2D\,\,$ electron
spectrum (e.g., HTSC) exhibit a NFL behavior in some temperature interval,
although the collective excitation spectrum is quite different from that in
the theory \cite{Var}. A similar situation takes place for $D=3$ provided
that nesting features of the Fermi surface are present$\,$ or spin-wave
spectrum possesses reduced dimensionality (Ce- and U-based systems). We
analyze also peculiar incoherent (``non-quasiparticle'') contributions to
electronic density of states and thermodynamic and transport properties,
which are not described by the standard Fermi-liquid theory.

In Sect.2 we calculate the electron Green's function for a conducting
antiferromagnet in the framework of the $s-d(f)$ exchange model. We
investigate various contributions to the electron self-energy and density of
states. In Sect.3 we calculate the electronic specific heat and transport
relaxation rate owing to the electron-magnon interaction. We analyze also
the incoherent contributions to transport properties, which are connected
with impurity scattering. A consistent quasiclassical perturbation theory is
discussed in Appendix A. A simple scaling consideration is performed in
Appendix B. Some of preliminary results of the paper were briefly presented
in Ref.\cite{IK95}.

\section{Peculiarities of electron spectrum: perturbation theory}

The peculiarities of spectrum and damping of quasiparticles near the Fermi
level are due to the interaction with low-energy collective excitations,
either well-defined or of dissipative nature (phonons, zero sound,
paramagnons etc.). Migdal \cite{Mig} proved for $D=3\,$ in a general form
that the corresponding non-analytic contributions to the self-energy $\Sigma
(E)\,$are of order of $E^3$ln$E$, which results in $T^3$ln$T$-terms in
electronic specific heat \cite{CP1}. The Fermi-liquid behavior might seem to
take place for AFM metals since in the long-wavelength limit ($q\rightarrow
0 $) the electron-magnon interaction is equivalent to the interaction with
acoustical phonons (the spectrum of the Bose excitations is linear and the
scattering amplitude is proportional to $q^{1/2}$). However, in the case of
AFM spin waves there exists one more ``dangerous'' region $\mathbf{q}%
\rightarrow \mathbf{Q}$ ($\mathbf{Q}$ is the wavevector of the AFM
structure), where the magnon frequency $\omega _{\mathbf{q}}$ tends to zero
and the scattering amplitude diverges as $\omega _{\mathbf{q}}^{-1/2}$. At
very small $E $ such processes are forbidden because of the presence of the
AFM splitting in the electron spectrum. At the same time, at not too small $%
E $ one may expect that these processes lead to stronger singularities. Thus
the Fermi-liquid picture may become violated in this energy region.

To investigate effects of interaction of current carriers with local moments
we use the Hamiltonian of the $s-d(f)$ exchange model
\begin{equation}
H=\sum_{\mathbf{k}\sigma }t_{\mathbf{k}}c_{\mathbf{k}\sigma }^{\dagger }c_{%
\mathbf{k}\sigma }-I\sum_{\mathbf{qk}}\sum_{\alpha \beta }\mathbf{S_q}c_{%
\mathbf{k}\alpha }^{\dagger }\mbox {\boldmath $\sigma $}_{\alpha \beta }c_{%
\mathbf{k-q}\beta }+\sum_{\mathbf{q}}J_{\mathbf{q}}\mathbf{S}_{\mathbf{q}}%
\mathbf{S}_{-\mathbf{q}}  \label{H}
\end{equation}
where $c_{\mathbf{k}\sigma }^{\dagger }$, $c_{\mathbf{k}\sigma }$ and $%
\mathbf{S}_{\mathbf{q}}$ are operators for conduction electrons and
localized spins in the quasimomentum representation, the electron spectrum $%
t_{\mathbf{k}}$ is referred to the Fermi level, $I$ is the $s-d(f)$ exchange
parameter, $\mathbf{\sigma }$ are the Pauli matrices. We consider an
antiferromagnet which has the spiral structure along the $x$-axis with the
wavevector \textbf{Q }
\[
\langle S_i^z\rangle =S\cos \mathbf{QR}_i,\,\langle \,S_i^y\rangle =S\sin
\mathbf{QR}_i,\,\langle S_i^x\rangle =0
\]
It is convenient to introduce the local coordinate system
\begin{eqnarray*}
S_i^z &=&\hat S_i^z\cos \mathbf{QR}_i-\hat S_i^y\sin \mathbf{QR}_i, \\
\,S_i^y &=&\hat S_i^y\cos \mathbf{QR}_i+\hat S_i^z\sin \mathbf{QR}%
_i,\,S_i^x=\hat S_i^x
\end{eqnarray*}
Further one can pass from spin operators $\mathbf{\hat S}_i$ to the spin
deviation operators $b_i^{\dagger },b_i$ and, by the canonical
transformation $b_{\mathbf{q}}^{\dagger }=u_{\mathbf{q}}\beta _{\mathbf{q}%
}^{\dagger }-v_{\mathbf{q}}\beta _{-\mathbf{q}},$ to the magnon operators.
Hereafter we consider for simplicity a two-sublattice AFM (2\textbf{Q }is
equal to a reciprocal lattice vector, so that $\cos ^2\mathbf{QR}_i=1,\,\sin
^2\mathbf{QR}_i=0 $). Then the Bogoliubov transformation coefficients and
the magnon frequency are given by
\begin{eqnarray}
u_{\mathbf{q}}^2 &=&1+v_{\mathbf{q}}^2=\frac 12[1+\overline{S}(J_{\mathbf{q+Q%
}}+J_{\mathbf{q}}-2J_{\mathbf{Q}})/\omega _{\mathbf{q}}] \\
\,\,\,\omega _{\mathbf{q}} &=&2\overline{S}(J_{\mathbf{Q-q}}-J_{\mathbf{Q}%
})^{1/2}(J_{\mathbf{q}}-J_{\mathbf{Q}})^{1/2}  \nonumber
\end{eqnarray}
so that $u_{\mathbf{q}}\mp v_{\mathbf{q}}\cong u_{\mathbf{q+Q}}\pm v_{%
\mathbf{q+Q}}\propto \omega _{\mathbf{q}}{}^{\pm 1/2}$at $q\rightarrow 0.$

Further we use at concrete calculations simple results of the usual
perturbation theory in $I$. Defining the self-energies by the perturbation
expansion 
\begin{equation}
G_{\mathbf{k}\sigma }(E)=\langle \langle c_{\mathbf{k}\sigma }|c_{\mathbf{k}%
\sigma }^{\dagger }\rangle \rangle _E=[E-t_{\mathbf{k}}-\sum_n\Sigma _{%
\mathbf{k}}^{(n)}(E)]^{-1}
\end{equation}
we derive for the contributions which contain the Fermi distribution
functions $n_{\mathbf{k}}=f(t_{\mathbf{k}})$ 
\begin{equation}
\Sigma _{\mathbf{k}}^{(2)}(E)=I^2\overline{S}\sum_{\mathbf{q}}(u_{\mathbf{q}%
}-v_{\mathbf{q}})^2\left( \frac{1-n_{\mathbf{k+q}}+N_{\mathbf{q}}}{E-t_{%
\mathbf{k+q}}-\omega _{\mathbf{q}}}+\frac{n_{\mathbf{k+q}}+N_{\mathbf{q}}}{%
E-t_{\mathbf{k+q}}+\omega _{\mathbf{q}}}\right)  \label{S2}
\end{equation}
\begin{equation}
\Sigma _{\mathbf{k}}^{(3)}(E)=I^3\overline{S}^2\sum_{\mathbf{q}}\left( \frac{%
1-n_{\mathbf{k+q}}+N_{\mathbf{q}}}{E-t_{\mathbf{k+q}}-\omega _{\mathbf{q}}}-%
\frac{n_{\mathbf{k+q}}+N_{\mathbf{q}}}{E-t_{\mathbf{k+q}}+\omega _{\mathbf{q}%
}}\right) \left( \frac 1{E-t_{\mathbf{k-Q}}}-\frac 1{t_{\mathbf{k+q}}-t_{%
\mathbf{k+q-Q}}}\right)  \label{S3}
\end{equation}
\begin{equation}
\Sigma _{\mathbf{k}}^{(4)}(E)=I^4\overline{S}^3\sum_{\mathbf{q}}(u_{\mathbf{q%
}}+v_{\mathbf{q}})^2\left( \frac 1{E-t_{\mathbf{k-Q}}}-\frac 1{E-t_{\mathbf{%
k+q-Q}}}\right) ^2\left( \frac{1-n_{\mathbf{k+q}}+N_{\mathbf{q}}}{E-t_{%
\mathbf{k+q}}-\omega _{\mathbf{q}}}+\frac{n_{\mathbf{k+q}}+N_{\mathbf{q}}}{%
E-t_{\mathbf{k+q}}+\omega _{\mathbf{q}}}\right)  \label{S4}
\end{equation}
where $N_{\mathbf{q}}=N_B(\omega _{\mathbf{q}})$ is the Bose function.

Non-analytic contributions to the self-energies at $E\rightarrow 0,$ $T=0$
originate from spin waves with small $q$ and $|\mathbf{q}-\mathbf{Q|}$.
Because of $q$-dependence of interaction matrix elements ($(u_{\mathbf{q}%
}\mp v_{\mathbf{q}})^2\propto q,(u_{\mathbf{q}}\pm v_{\mathbf{q}})^2\propto
q^{-1}$at $q\rightarrow 0 $ and $\mathbf{q}\rightarrow \mathbf{Q}$
respectively), the intersubband contributions ($\mathbf{q}\rightarrow 
\mathbf{Q}$) are, generally speaking, more singular than intrasubband ones ( 
$q\rightarrow 0$). However, owing to quasimomentum and energy conservation
laws, the intersubband transitions are possible at $|\mathbf{q-Q}|>q_0\sim
\Delta /v_F$ ($\Delta =2I\overline{S}$ is the antiferromagnetic splitting, $%
\overline{S}$ is the sublattice magnetization, $v_F$ is the electron
velocity at the Fermi level). Therefore, when using simple perturbation
expressions, one has to bear in mind that the singular intersubband
transition contributions should be cut at $|E|,T\sim T^{*}$ where 
\begin{equation}
T^{*}=cq_0\sim T_N\Delta /v_F  \label{T*}
\end{equation}
with $c$ being the magnon velocity. A more general perturbation theory is
considered in Appendix A.

It should be noted that, despite absence of long-range order at finite
temperatures, the results (\ref{S2})-(\ref{S4}) are valid also in the $2D$
case up to $T\sim J$, $\overline{S}$ being replaced by square root of the
Ornstein-Cernike peak intensity in the pair correlation function \cite{IKCM1}%
. We have also to replace $T_N\rightarrow JS^2$ in (\ref{T*}). A similar
situation occurs for frustrated magnetic systems with suppressed long-range
order. In particular, one can think that the consideration of electron-spin
interactions, that is based on the spin-wave picture, is qualitatively
applicable to Y$_{1-x}$Sc$_x$Mn$_2,$ despite this is not an antiferromagnet,
but a spin liquid with strong short-range AFM order.

The correction to the density of states owing to $s-d(f)$ interaction reads 
\begin{equation}
\delta N(E)=-\sum_{\mathbf{k}i\sigma }[\frac 1\pi \mathrm{Im}\Sigma _{%
\mathbf{k}}^{(i)}(E)\mathbf{/}(E-t_{\mathbf{k}})^2+\mathrm{Re}\Sigma _{%
\mathbf{k}}^{(i)}(E)\delta ^{\prime }(E-t_{\mathbf{k}})]  \label{N(E)}
\end{equation}
The first term in (\ref{N(E)}) corresponds to incoherent (non-quasiparticle)
contribution, and the second one describes the renormalization of
quasiparticle spectrum.

The third-order contribution (\ref{S3}) describes the Kondo effect in the
AFM state \cite{IKCM1,IKZ}. It should be noted that an account of spin
dynamics is important at treating this contribution, and its neglect leads
to incorrect results: the transition to the usual Kondo behavior Im$\Sigma
(E)\propto \ln |E|$ discussed in Ref.\cite{arai} takes place in fact only at 
\[
|E|\gg \overline{\omega }=\omega (2k_F)\simeq 2ck_F. 
\]

Unlike $\Sigma _{\mathbf{k}}^{(2)}(E)$, $\Sigma _{\mathbf{k}}^{(4)}(E)$ does
not contain ``dangerous'' divergences since the factor $(u_{\mathbf{q}}+v_{%
\mathbf{q}})^2$ is singular at $q\rightarrow 0$ (rather than at $|\mathbf{{%
q-Q}|}\rightarrow 0$) and the next factor is proportional to $q^2$. Thus
this term results in unimportant renormalizations of $\Sigma _{\mathbf{k}%
}^{(2)}(E).$ Summation of higher-order correction within a scaling approach
is presented in Appendix B.

Averaging $\Sigma _{\mathbf{k}}^{(2)}(E)$ over the Fermi surface $t_{\mathbf{%
k}}=0,$ we obtain for the intrasubband contribution at $T\ll |E|$ in the $3D$
case 
\begin{equation}
\left\{ 
\begin{array}{c}
\mathrm{Re} \\ 
\mathrm{Im}
\end{array}
\right\} \Sigma ^{(2)}(E)=\frac{2I^2\rho }{3\overline{\omega }^2(J_0-J_{%
\mathbf{Q}})}E^3\times \left\{ 
\begin{array}{c}
\ln |E/\overline{\omega }| \\ 
-(\pi /2)\mathrm{sgn}E
\end{array}
\right.  \label{sintr}
\end{equation}
Thus, after analytical continuation, the contributions to Im$\Sigma (E),$
proportional to $E^2|E|,$ result in corrections of the form $\delta $Re$%
\Sigma (E)\propto E^3$ln$|E|$, which is in agreement with the microscopic
Fermi-liquid theory \cite{Mig}. Then the second term in (\ref{N(E)}) yields
the contributions of the form $\delta N(E)\propto E^2$ln$|E|$. For $D=2,$ Im$%
\Sigma ^{(2)}(E)$ is proportional to $E^2$ and does not result in occurrence
of non-analytic terms in Re$\Sigma (E)$ and $N(E)$. Note that in a $2D$
paramagnet electron-electron scattering results in the contributions Im$%
\Sigma (E)\propto E^2\ln |E|$, and in $T^2\ln T$-terms in resistivity \cite
{Hod}.

As for the ``Kondo'' (third-order) term (\ref{S3}), picking out the most
singular contribution yields 
\begin{equation}
\delta \Sigma _{\mathbf{k}}^{(3)}(E)=-\frac{2I^3S^2}{E-t_{\mathbf{k-Q}}}%
\left\langle \frac 1{t_{\mathbf{k}^{\prime }\mathbf{-Q}}}\right\rangle _{t_{%
\mathbf{k}^{\prime }}=0}\sum_{\mathbf{k}^{\prime }}\frac{n_{\mathbf{k}%
^{\prime }}(E-t_{\mathbf{k}^{\prime }})^2}{(E-t_{\mathbf{k}^{\prime
}})^2-\omega _{\mathbf{k-k}^{\prime }}^2}
\end{equation}
Thus the singularity in this term is by a factor of $|E|$ weaker in
comparison with the intrasubband contribution to $\Sigma ^{(2)}(E)$ (note
that the corresponding results of Ref.\cite{IKZ,IK95} are not correct since
not all $q$-dependent factors were taken into account).

Now we investigate the intersubband contributions. Averaging (\ref{S2}) in $%
\mathbf{k}$\textbf{\ }over the Fermi surface $t_{\mathbf{k}}=E_F=0$ we
obtain 
\begin{equation}
\mathrm{Im}\Sigma (E)=-2\rho ^{-1}\sum_{\mathbf{q\simeq Q,\ }T^{*}\leq
\omega _{\mathbf{q}}<|E|}\frac{\lambda _{\mathbf{q}}}{\omega _{\mathbf{q}}}
\label{IMS}
\end{equation}
with $\rho =\sum_{\mathbf{k}}\delta (t_{\mathbf{k}})$ the bare density of
states at the Fermi level, 
\begin{equation}
\lambda _{\mathbf{q}}=2\pi I^2\overline{S}^2(J_0-J_{\mathbf{Q}})\sum_{%
\mathbf{k}}\delta (t_{\mathbf{k}})\delta (t_{\mathbf{k+q}})  \label{lam}
\end{equation}
In the general $3D$ case we have Im$\Sigma (E)\propto E^2$. For $D=2$ we
derive 
\begin{equation}
\left\{ 
\begin{array}{c}
\mathrm{Re} \\ 
\mathrm{Im}
\end{array}
\right\} \Sigma ^{(2)}(E)=\frac 2{\pi ^2\rho c^2}\lambda _{\mathbf{Q}%
}E\times \left\{ 
\begin{array}{c}
\ln |E/\overline{\omega }| \\ 
-(\pi /2)\mathrm{sgn}E
\end{array}
\right.
\end{equation}
so that Im$\Sigma (E)$ is linear in $|E|.$ The residue of the electron
Green's function 
\begin{equation}
Z=\left( \left. 1-\frac{\partial \mathrm{Re}\Sigma (E)}{\partial E}\right|
_{E=E_F}\right) ^{-1}  \label{z}
\end{equation}
yields the renormalization of the effective mass 
\begin{equation}
m^{*}/m=1/Z\sim \ln \left( \overline{\omega }/\omega ^{*}\right) .
\end{equation}
The second term of (\ref{N(E)}) yields at $|E|>T^{*}$%
\begin{equation}
\delta N(E)=-\frac 4{\pi ^2c^2}\lambda _{\mathbf{Q}}\ln |E|  \label{ncoh}
\end{equation}

Consider the peculiar $3D$ case where the electron spectrum satisfies
approximately the ``nesting'' condition $t_{\mathbf{k}}=-t_{\mathbf{k+Q}}$
in a significant part of the Brillouin zone (however, the system is still
metallic since the gap does not cover the whole Fermi surface). Such a
situation is typical for itinerant-electron AFM systems since onset of AFM
ordering is connected with the nesting. Besides that, for localized-moment
metallic magnets, that are described by the $s-f$ model, the value of $%
\mathbf{Q}$\textbf{\ }is also often determined by the nesting condition \cite
{Dz}. As discussed in the Introduction, such a situation can be also assumed
for some anomalous $f$-systems.

In the case under consideration the electron spectrum near the Fermi surface
is strongly influenced by the AFM gap, so that we have to use the ``exact''
spectrum (\ref{specr}) and replace in (\ref{lam}) 
\[
\delta (t_{\mathbf{k}})\delta (t_{\mathbf{k+q}})\rightarrow \delta (E_{%
\mathbf{k}1})\delta (E_{\mathbf{k+q}2}) 
\]
Then we have in some $\mathbf{q}$\textbf{-}region (which is determined not
only by $I$ but also by characteristics of the Fermi surface) $\lambda _{%
\mathbf{q}}\propto 1/|\mathbf{{q-Q}|\,.}$ Thus the effective dimensionality
in the integrals is reduced by unity, and the energy and temperature
dependences become similar to those in the $2D$ case.

For the $2D$ ``nested'' antiferromagnet, the perturbation theory damping is
very large, 
\begin{equation}
\mathrm{Im}\Sigma (E)\propto \ln |E|.
\end{equation}
Replacing the denominator in (\ref{S2}) by the exact electron Green's
function and making the ansatz Im$\Sigma (E)\propto |E|^a$ we can estimate
the damping in the second-order self-consistent approximation as 
\begin{equation}
\mathrm{Im}\Sigma (E)\propto |E|^{1/2}.
\end{equation}
Thus one has to expect in this case a strongly non-Fermi-liquid behavior at
not too small $|E|$. Note that the situation is different from the power-law
non-analycity in the Anderson model at very small $|E|$, which is an
artifact of the NCA approximation \cite{Cox}.

The damping (\ref{IMS}) becomes stronger also in the case of frustrations in
the localized spin subsystems where the \textbf{q}-dependence of magnon
frequency 
\begin{equation}
\omega _{\mathbf{q}}^2=c_x^2q_x^2+c_y^2q_y^2+c_z^2q_z^2+\Phi ^{(4)}(\mathbf{q%
})+...
\end{equation}
($\Phi ^{(4)}(\mathbf{q})$ is a quartic form of $q_x,q_y,q_z$) becomes
anomalous. As discussed above, such a situation is typical for systems
demonstrating NFL behavior. Instabilities of magnetic structures with
competing exchange interactions are often accompanied by softening of magnon
spectrum. Usually this takes place in one direction, i.e. $c_z\ll c_x,c_y$
near the instability point. In some peculiar models, the softening can occur
in two or even three directions. In all the cases, the energy dependences of 
$\Sigma $ are changed. We have 
\begin{equation}
\mathrm{Im}\Sigma (E)\simeq -2\rho ^{-1}\lambda _{\mathbf{Q}%
}\int_{T^{*}}^{|E|}\frac{d\omega }\omega g(\omega )\propto |E|^\alpha
\end{equation}
where 
\begin{equation}
g(\omega )=\sum_{\mathbf{q\simeq Q}}\delta (\omega -\omega _{\mathbf{q}%
})\propto \omega ^\alpha ,\,\omega \rightarrow 0
\end{equation}
Provided that $c_z\ll c_x,c_y,$ one has to take into account quartic terms,
and we obtain after passing to cylindrical coordinates $\alpha =3/2.$ In the
opposite case $c_z\gg c_x,c_y$ we get in a similar way $\alpha =1.$ For $%
c_x,c_y,c_z\rightarrow 0$ we derive $\alpha =1/2.$ In the 2D case we have $%
\alpha =1/2$ at $c_x\ll c_y$ and $\alpha =0$ (Im$\Sigma (E)\propto \ln
|E/T^{*}|$) at $c_x,c_y\rightarrow 0.$ Thus we can explain violations of the
Fermi-liquid picture by peculiarities of not only electron, but also magnon
spectrum. The frustration problem for an itinerant AFM was considered in Ref.%
\cite{Lac}.

Now we treat the incoherent contribution to $N(E)$ (first term in (\ref{N(E)}%
)). We have at $T=0$ 
\begin{eqnarray}
\delta N_{incoh}(E) &\simeq &I^2S\sum_{\mathbf{q}\sigma }(u_{\mathbf{q}}-v_{%
\mathbf{q}})^2\sum_{\mathbf{k}}\frac{\mathcal{P}}{(E-t_{\mathbf{k}})^2} 
\nonumber \\
&&\ \ \ \ \ \times \ \ [(1-n_{\mathbf{k+q}})\delta (E-t_{\mathbf{k+q}%
}-\omega _{\mathbf{q}})+n_{\mathbf{k+q}}\delta (E-t_{\mathbf{k+q}}+\omega _{%
\mathbf{q}})],
\end{eqnarray}
where $\mathcal{P}$ stands for the principal value of the integral. After a
little manipulation we obtain 
\begin{equation}
\delta N_{incoh}(E)\simeq I^2S\sum_{\mathbf{q,}\omega _{\mathbf{q}}<|E|}(u_{%
\mathbf{q}}-v_{\mathbf{q}})^2\mathcal{P}\sum_{\mathbf{k}\sigma }\frac{\delta
(|E|-t_{\mathbf{k+q}}-\omega _{\mathbf{q}})}{(E-t_{\mathbf{k}})^2}
\label{incoh}
\end{equation}
In the sum over $\mathbf{k}$ we can neglect $|E|$ and $\omega _{\mathbf{q}}$
in comparison with $v_Fq.$ Main contribution comes from the intersubband
transitions, 
\begin{equation}
\delta N_{incoh}(E)\simeq I^2S\sum_{\mathbf{q\rightarrow Q,}\omega _{\mathbf{%
q}}<|E|}\frac{2S(J_0-J_{\mathbf{Q}})}{\omega _{\mathbf{q}}}\mathcal{P}\sum_{%
\mathbf{k}\sigma }\frac{\delta (t_{\mathbf{k}})}{(t_{\mathbf{k+Q}}-t_{%
\mathbf{k}})^2}\propto \int_0^{|E|}\frac{d\omega }\omega g(\omega )\propto
|E|^\alpha
\end{equation}
This contribution can have any sign depending on the dispersion law $t_{%
\mathbf{k}}$. For the general electron and magnon spectra we have $\delta
N_{incoh}(E)\propto E^2$ for $D=3$ and $\delta N_{incoh}(E)\propto |E|$ for $%
D=2.$ Unlike the quasiparticle contribution (\ref{ncoh}), the incoherent one
is not cut at $|E|\simeq T^{*}$ since the conservation laws do not work for
virtual magnons. In the case of ``frustrated'' magnon spectrum the values of 
$\alpha $ are given above. In the nesting situation the additional divergent
factor $1/|\mathbf{q-Q}|$ occurs which is, however, cut at $q=q_0.$ This
leads to that, at $|E|>T^{*},$ $\delta N_{incoh}(E)\propto |E|$ for $D=3$
and $\delta N_{incoh}(E)\propto -\ln |E|$ for $D=2.$ Due to the factor $(u_{%
\mathbf{q}}-v_{\mathbf{q}})^2,$ the contribution from the region of small $q$
in (\ref{incoh}) contains higher powers of $E,$ despite the singularity in
the denominator. The third-order ``Kondo'' contribution to $\delta
N_{incoh}(E)$ is small owing to cancellation of intra- and intersubband
transitions in (\ref{S3}).The ``incoherent'' contributions may be in
principle observed in tunneling experiments.

Now we treat for comparison the case of a metallic ferromagnets with the
parabolic dispersion law of magnons ($\mathbf{Q}=0$). We have for a given
spin projection 
\begin{equation}
\Sigma _{\mathbf{k\pm }}^{(2)}(E)=2I^2\overline{S}\sum_{\mathbf{q}}\frac{%
f(\pm t_{\mathbf{k+q}}+IS)+N_{\mathbf{q}}}{E-t_{\mathbf{k+q}}\mp IS\pm
\omega _{\mathbf{q}}}
\end{equation}
This yields for $D=3$ the one-sided singular contributions 
\begin{equation}
\mathrm{Im}\Sigma _\sigma (E)\propto \theta (\sigma
E)|E|^{3/2},\,\,\,|E|>T^{*}\sim (\Delta /v_F)^2T_C,
\end{equation}
the crossover energy scale being considerably smaller than in the AFM case
(see \cite{Aus,IKCM2}; these papers treat also the quasiparticle damping at
small $|E|$ due to electron-magnon scattering, which occurs in the second
order in $1/2S$ and turns out to be small). Then we obtain $\delta N_\sigma
(E)\propto -\ln |E|$ for $|E|>T^{*}$. At the same time, the incoherent
contribution, which survives up to $E=0$, has the form $\delta N_\sigma
(E)\propto \theta (\sigma E)|E|^{3/2}$ and can be picked up in the
half-metallic case \cite{Edw,Aus}.

The situation in a $2D$ ferromagnet is similar to that in the
above-discussed ``nested'' $2D$ antiferromagnet: the damping in the
perturbation theory is large, Im$\Sigma (E)\propto |E|$ $^{1/2}$, this
result being valid in the self-consistent approximation too.

\section{Thermodynamic and transport properties}

To calculate the electronic specific heat $C(T)$ we use the thermodynamic
identity 
\begin{equation}
\left( \frac{\partial n}{\partial T}\right) _\mu =\left( \frac{\partial 
\mathcal{S}}{\partial \mu }\right) _T
\end{equation}
with $n$ the number of particles, $\mathcal{S}$ the entropy, $\mu $ the
chemical potential. Taking into account the expression 
\begin{equation}
n=\int_{-\infty }^\infty dEf(E)N(E)  \label{ne}
\end{equation}
and integrating by part, we obtain from the second term of (\ref{N(E)}) 
\[
\delta n=\frac \partial {\partial \mu }\Phi ,\,\delta C(T)=\delta \mathcal{S}%
(T)=\frac \partial {\partial T}\Phi 
\]
where, to lowest order in $I$ 
\begin{equation}
\Phi =-2I^2\overline{S}\sum_{\mathbf{k,q}}(u_{\mathbf{q}}-v_{\mathbf{q}})^2%
\frac{\omega _{\mathbf{q}}n_{\mathbf{k}}(1-n_{\mathbf{k+q}})}{(t_{\mathbf{k}%
}-t_{\mathbf{k+q}})^2-\omega _{\mathbf{q}}^2}.  \label{Fi}
\end{equation}
It should be noted that the same result for $C(T)$ can be derived by
calculating the transverse-fluctuation contribution to the interaction
Hamiltonian 
\begin{equation}
\delta \langle H_{sd}\rangle =-I(2S)^{1/2}\langle b_{\mathbf{q}}^{\dagger
}(c_{\mathbf{k}\downarrow }^{\dagger }c_{\mathbf{k+q}\uparrow }+c_{\mathbf{k}%
\uparrow }^{\dagger }c_{\mathbf{k+q}\downarrow })\rangle =-2\Phi  \label{Hsd}
\end{equation}
(the last equality in (\ref{Hsd}) is obtained from the spectral
representation for the corresponding Green's function, cf. Ref.\cite{IEnt})
and using the Hellman-Feynman theorem for the free energy, $\partial 
\mathcal{F}/\partial I=\langle \partial H/\partial I\rangle $.

Using the identity 
\begin{eqnarray}
&&\ \frac \partial {\partial T}\int \int dEdE^{\prime }f(E)[1-f(E^{\prime
})]F(E-E^{\prime })  \nonumber \\
\ &=&\frac 1T\int \int dEdE^{\prime }E\frac{\partial f(E)}{\partial E}\frac{%
\partial f(E^{\prime })}{\partial E^{\prime }}\int_{E^{\prime
}-E}^{E^{\prime }+E}dxF(x)
\end{eqnarray}
we obtain for the intrasubband ($q\rightarrow 0$) contribution to (\ref{Fi})
at $D=3$%
\begin{equation}
\delta C_{intra}(T)=\frac{74}{135}\frac{\pi ^4I^2\rho ^2}{(J_0-J_{\mathbf{Q}%
})\bar \omega ^2}T^3\ln \frac{\bar \omega }T
\end{equation}
The term, proportional to $T^3\ln (T/T_{sf})$ ($T_{sf}$ is a characteristic
spin-fluctuation energy, in our case $T_{sf}\sim |J|$) was derived earlier
within the Fermi-liquid theory \cite{CP1}. Note that the $T^3\ln T$%
-corrections were not obtained in the spin-fluctuation theory by Moriya et
al \cite{Morb} since only fluctuations with $\mathbf{q\simeq Q}$ were taken
into account.

The intersubband contribution to specific heat is transformed as

\begin{equation}
\delta C_{inter}(T)=\frac 83\pi T\sum_{\mathbf{q\simeq Q,}T\leq \omega _{%
\mathbf{q}}}\lambda _{\mathbf{q}}/\omega _{\mathbf{q}}^2  \label{intc}
\end{equation}
In the $2D$ or ``nesting'' $3D$ situation the integral is logarithmically
divergent at $\mathbf{q\rightarrow Q}$\textbf{$,\ $}and the divergence is
cut at $\omega _{\mathbf{q}}\simeq \max (T,T^{*}),$ so that we obtain the $%
T\ln T$-dependence of specific heat. For $D=2$ we have 
\begin{equation}
\delta C_{inter}(T)=\frac{4\Omega _0}{3c^2}\lambda _{\mathbf{Q}}T\ln \frac{%
\bar \omega }{\max (T,T^{*})}  \label{cinter}
\end{equation}
Since the integral in (\ref{intc}) is determined by the magnon spectrum
only, the result (\ref{cinter}) holds also in the case the frustrated ($2D$%
-like) magnon spectrum. This contribution may explain the anomalous
dependences $C(T)$ in a number of rare-earth and actinide systems (see the
Introduction), which are observed in some restricted temperature intervals.
At $T<T^{*}$ we have an appreciable logarithmic enhancement of the
electronic specific heat.

For $T>T^{*}$ the $T\ln T$-term is present also in specific heat of a
ferromagnet, both for the case of weak itinerant-electron ferromagnetism 
\cite{DzK} and in the regime of local moments \cite{IKZ,IKCM2}. However, we
shall see below that the ferromagnets do not exhibit an important property
of the MFL state - the $T$-linear resistivity.

In the model accepted, the non-analytic contributions to magnetic
susceptibility should mutually cancel, as well as for electron-phonon
interaction \cite{prange}. However, such contributions may occur in the
presence of relativistic interactions (e.g., for heavy actinide atoms). This
effects may be responsible for anomalous $T$-dependences of $\chi $ in the $%
4f$- and $5f$-systems \cite{Map}.

The first term in (\ref{N(E)}) (i.e. the branch cut of the self-energy)
yields the incoherent (non-quasiparticle) $T$-linear contribution to
specific heat (cf. the consideration for a ferromagnet in Ref.\cite{IKCM2}),
which is owing to the temperature dependence of $N(E)$ and is not described
by the Fermi liquid theory. After substituting this term into (\ref{ne}) we
obtain 
\begin{equation}
\frac \partial {\partial T}(\delta n)_{incoh}=2I^2\overline{S}\sum_{\mathbf{%
k,q},\alpha =\pm }(u_{\mathbf{q}}-v_{\mathbf{q}})^2\frac{f(t_{\mathbf{k+q}%
}-\alpha \omega _{\mathbf{q}})}{(t_{\mathbf{k+q}}-t_{\mathbf{k}}-\alpha
\omega _{\mathbf{q}})^2}\frac \partial {\partial T}n_{\mathbf{k+q}}
\label{ninc}
\end{equation}
At low temperatures we have 
\[
f(t_{\mathbf{k+q}}+\omega _{\mathbf{q}})\rightarrow 0,\,\,\,f(t_{\mathbf{k+q}%
}-\omega _{\mathbf{q}})\rightarrow 1 
\]
and we derive 
\begin{equation}
\delta C_{incoh}(T)=\frac 23\pi ^2I^2\overline{S}\rho T\sum_{\mathbf{q}%
}\left\langle \left( \frac{u_{\mathbf{q}}-v_{\mathbf{q}}}{t_{\mathbf{k+q}%
}-t_{\mathbf{k}}}\right) ^2\right\rangle _{t_{\mathbf{k}}=0}  \label{cinc}
\end{equation}
Note that the contribution (\ref{cinc}) can be also obtained by direct
differentiating in temperature the total electronic energy 
\begin{equation}
\mathcal{E}=\int_{-\infty }^\infty dEEf(E)N(E)
\end{equation}

Now we discuss transport properties. To second order in $I$, using the Kubo
formula \cite{Nak} we obtain for the inverse transport relaxation time 
\begin{equation}
1/\tau =\pi I^2\bar S\sum_{\mathbf{kq}}(u_{\mathbf{q}}-v_{\mathbf{q}})^2(%
\mathbf{v_{k+q}-v}_{\mathbf{k}})^2\delta (t_{\mathbf{k+q}}-t_{\mathbf{k}%
}-\omega _{\mathbf{q}})\delta (t_{\mathbf{k}})/\sum_{\mathbf{k}}\mathbf{v}_{%
\mathbf{k}}^2\delta (t_{\mathbf{k}})  \label{tau}
\end{equation}
with $\mathbf{v}_{\mathbf{k}}=\partial t_{\mathbf{k}}/\partial \mathbf{k.\ }$%
Picking out the intersubband contribution ($\mathbf{q}\simeq \mathbf{Q}$) we
obtain after standard transformations 
\begin{equation}
\frac 1\tau =\frac{\langle (\mathbf{v_{k+Q}-v}_{\mathbf{k}})^2\rangle _{t_{%
\mathbf{k}}=0}}{v_F^2\rho }\sum_{\mathbf{q\simeq Q}}\lambda _{\mathbf{q}%
}\left( -\frac{\partial N_{\mathbf{q}}}{\partial \omega _{\mathbf{q}}}\right)
\label{taul}
\end{equation}
For $D=3$ we have the quadratic temperature dependence of spin-wave
resistivity, 
\begin{equation}
R(T)\propto (T/T_N)^2.
\end{equation}
This dependence was obtained earlier within an itinerant model \cite{Ueda}.
Note that this contribution dominates at not too low temperatures over the
intrasubband contribution (the latter is analogous to the electron-phonon
scattering one and is proportional to $T^5\,$\cite{Ros})$.$

For the $2D$ magnon spectrum (or ``nested'' $3D$) situation one obtains 
\begin{equation}
R(T)\propto T\ln (1-\exp (-T^{*}/T))\simeq T\ln (T/T^{*})  \label{rtlnt}
\end{equation}
Thus in our model, unlike \cite{Var}, the linear dependence of Im$\Sigma (E)$
results in $T\ln T$ rather than $T$-linear behavior of the resistivity
because of the lower-limit divergence of the integral with the Bose
function. However, the deviation from the linear law is hardly important
from the experimental point of view. As for concrete experimental data, the
systems CePd$_2$Si$_2$ and CeNi$_2$Ge$_2$ \cite{Jul} demonstrate under
pressure anomalous temperature dependence $\rho (T)\sim T^\mu ,\mu =$ $%
1.2\div 1.5.$ The data of Ref.\cite{Steg} on CeNi$_2$Ge$_2$ yield for the
resistivity exponent $\mu =3/2.$

For a ferromagnet, the spin-wave resistivity at $T>T^{*}$ is proportional to 
$T^2$ for $D=3$ (and $T^{3/2}$ for $D=2$) because of the factor $(\mathbf{v}%
_{\mathbf{k}}-\mathbf{v}_{\mathbf{k+q}})^2$ in (\ref{tau}). (However, extra
powers of $q $ are absent for the scattering between spin subbands, which
yields the $T\ln T$-term in resistivity of ferromagnetic alloys \cite{MFK}.)
A similar situation takes place in the case of ``flat'' regions of the Fermi
surface in AFM. This may explain absence of $T$-linear resistivity in some
above-discussed rare-earth and actinide systems which demonstrate $T\ln T$%
-corrections to specific heat.

Now we treat the impurity contributions to transport properties in the
presence of potential scattering (in the case of a ferromagnet they were
considered in \cite{IKT}). To second order in impurity potential $V$ we
derive 
\begin{equation}
\langle \langle c_{\mathbf{k}\sigma }|c_{\mathbf{k}^{\prime }\sigma
}^{\dagger }\rangle \rangle _E=\delta _{\mathbf{kk}^{\prime }}G_{\mathbf{k}%
\sigma }(E)+VG_{\mathbf{k}\sigma }(E)G_{\mathbf{k}^{\prime }\sigma
}(E)[1+V\sum_{\mathbf{p}}G_{\mathbf{p}\sigma }(E)]
\end{equation}
Neglecting vortex corrections and averaging over impurities we obtain for
the transport relaxation time 
\begin{equation}
\delta \tau _{imp}^{-1}(E)=-2V^2\mathrm{Im}\sum_{\mathbf{p}}G_{\mathbf{p}%
\sigma }(E)
\end{equation}
Thus the contributions under consideration are determined by the energy
dependence of $N(E)$ near the Fermi level. The correction to resistivity
reads 
\begin{equation}
\delta R_{imp}(T)/R^2=-\delta \sigma _{imp}(T)\propto -V^2\int dE(-\partial
f(E)/\partial E)\delta N(E)  \label{rimp}
\end{equation}
Note that the quasiparticle renormalization effects owing to $1-d\mathrm{Re}%
\Sigma (E)/dE=1/Z$ do not contribute impurity scattering since $\tau
\rightarrow \tau /Z$ and $v_F$\textbf{$\rightarrow v_FZ,$} so that the mean
free path is unrenormalized \cite{Mah}. At the same time, incoherent terms
in $N(E)$ yield $\delta R_{imp}(T)\propto T^2$ in the $3D$ case and $\delta
R_{imp}(T)\propto T$ in the $2D$ case up to lowest temperatures (in the
``nested'' $3D$ case, the $T$-linear term has lower cutoff, as well as the
``coherent'' contribution (\ref{rtlnt})). In the ``frustrated'' $3D$ case
with $\alpha =1/2$ we have $\delta R_{imp}(T)\propto T^{3/2}.$

The impurity contributions are important in ``dirty'' nearly AFM metals
where anomalous contributions to the temperature dependence of resistivity
can be both positive and negative. In particular, for the system U$_x$Y$%
_{1-x}$Pd$_3$ the experimental data \cite{Map} demonstrate the negative
contribution to resistivity, $\delta \rho (T)\sim -T^\mu ,$ $\mu =$ $1.1\div
1.4.$ In such cases the explanation of these terms by the spin-wave
renormalization of impurity scattering seems to be reasonable.

The correction to thermoelectric power, which is similar to (\ref{rimp}),
reads (cf.\cite{IKT}): 
\begin{equation}
\delta \mathcal{Q}(T)\propto \frac 1T\int dE(-\partial f(E)/\partial
E)E\delta N(E)  \label{qe}
\end{equation}
Besides that, an account of higher orders in impurity scattering leads to
the replacement of the impurity potential $V$ by the $T$-matrix. For the
point-like scattering the latter quantity is given by 
\begin{equation}
T(E)=\frac V{1-V\mathcal{R}(E)},\mathcal{R}(E)=\mathcal{P}\sum_{\mathbf{k}%
}G_{\mathbf{k}\sigma }(E).  \label{te}
\end{equation}
Expanding (\ref{te}) yields also the term 
\begin{equation}
\delta \mathcal{Q}(T)\propto \frac 1T\int dE(-\partial f(E)/\partial
E)E\delta \mathcal{R}(E)  \label{qte}
\end{equation}
with $\delta \mathcal{R}(E)$ being obtained by analytical continuation from $%
\delta N(E).$ Unlike the case of a ferromagnet where $\delta \mathcal{Q}%
(T)\propto T^{3/2}$ for $D=3$, the $I^2$-contribution to $\delta
N_{incoh}(E) $ in the AFM case is even in $E$ and does not contribute (\ref
{qe}). At the same time, in the $2D$ case, where $\delta \mathcal{R}%
(E)\propto E\ln |E|$, Eq.(\ref{qte}) yields 
\begin{equation}
\delta \mathcal{Q}(T)\propto T\ln (T/\overline{\omega })
\end{equation}
For the ``nested'' $3D$ case such contributions are present at $T>T^{*}$
only. In this connection, experimental data on the systems Y(Mn$_{1-x}$Al$_x$%
)$_2,$ Y$_{1-x}$Sc$_x$Mn$_2$ \cite{YMNR} are of interest which demonstrate
anomalous behavior of $\mathcal{Q}(T)$.

\section{Conclusions}

We have investigated peculiarities of electron spectrum and corresponding
anomalies of thermodynamic and transport properties in metallic
antiferromagnets with well-localized magnetic moments. The use of
perturbation theory in the electron-magnon interaction within the $s-d(f)$
exchange model seems to be a reasonable phenomenological approach for
highly-correlated electron systems, which takes into account the SU(2)
symmetry of exchange interactions. It is often used, e.g., in the
theoretical description of high-$T_c$ copper-oxide superconductors (see,
e.g., Ref.\cite{pines}). The electron spectrum $t_{\mathbf{k}}$ and
parameter $I $ may be considered as effective ones (including many-electron
renormalizations). Note that similar results for the electron-magnon
interaction effects may be obtained in the Hubbard model ($I\rightarrow U$,
cf.\cite{IKCM2,IEnt}).

We have demonstrated that, owing to intersubband scattering processes,
electronic properties of $2D$ and ``nested'' $3D$ metallic antiferromagnets
are close to those in the MFL picture \cite{Var} in a rather wide interval $%
T^{*}<T<J,$ the value of the crossover temperature being determined by the $%
s-f$ exchange parameter and characteristics of electron spectrum. In
contrast to \cite{Var}, no special assumptions about the spectrum of the
Bose excitations are used: in our model they are just spin waves with the
linear dispersion law. Unlike Refs.\cite{Kampf,Mor}, we need not to consider
the special case of the vicinity to AFM instability. Thus AFM ordering
itself, together with rather natural assumptions about a peculiar form of
the electron or magnon spectrum, may explain violations of the Fermi-liquid
picture which are observed in some rare-earth and actinide systems.

At $T<T^{*}\,$the MFL behavior is changed by the usual Fermi-liquid one,
although some non-quasiparticle contributions are present, which are
connected with the presence of local magnetic moments. These incoherent
contributions, which are beyond the Fermi liquid theory, play the crucial
role for half-metallic ferromagnets \cite{UFN}. In AFM metals they are not
so important and are hardly observable for perfect crystals. Nevertheless,
such contributions may be important for the temperature dependences of
transport properties in the ``dirty'' case (metals with impurities).

We have also analyzed the Kondo contributions to electronic properties in
the AFM state. In the case under consideration, they turn out to be strongly
suppressed by spin dynamics. Thus main role belongs to the second-order
corrections, and higher orders in $I$ are not important. Formally it is a
consequence of the divergence of the factors $(u_{\mathbf{q}}-v_{\mathbf{q}%
})^2$ at $\mathbf{q}\rightarrow \mathbf{Q}$; the Kondo terms do not contain
this divergence \cite{IKZ,kondo}. The situation should change with
increasing $|I| $ when renormalization of magnon frequencies becomes
important and summation of the higher orders is needed. Within a simple
scaling approach, such a problem was considered in Ref.\cite{kondo}. Thus,
the transition from ``usual'' magnets with well-defined local moments, which
are weakly coupled to conduction electrons, to the anomalous Kondo magnets
is accompanied by a reconstruction of the structure of perturbation theory
(different diagram sequences dominate in these two regimes). Therefore the
problem of the formulation of an unified picture of metallic magnetism
appears to be very complicated not only for itinerant $d$-electron magnets 
\cite{Morb,UFN}, but also for $f$-electron ones.

The research described was supported in part by Grant No.99-02-16279 from
the Russian Basic Research Foundation.

\appendix

\section{Self-energies in the $1/S$-expansion}

The electron spectrum in the AFM phase contains two split subbands. In the
mean-field approximation we have 
\begin{eqnarray}
E_{\mathbf{k}1,2} &=&\theta _{\mathbf{k}}\pm E_{\mathbf{k}}\,,E_{\mathbf{k}%
}=(\tau _{\mathbf{k}}^2+I^2\overline{S}^2)^{1/2},  \label{specr} \\
\theta _{\mathbf{k}} &=&\frac 12(t_{\mathbf{k}}+t_{\mathbf{k}+\mathbf{Q}%
}),\,\tau _{\mathbf{k}}=\frac 12(t_{\mathbf{k}}-t_{\mathbf{k}+\mathbf{Q\ }})
\end{eqnarray}
To calculate the fluctuation corrections in a consistent way, we have to
separate effects of transition within and between the AFM subbands by
including the AFM splitting in the zero-order approximation. Introducing
spinor operators $\Psi _{\mathbf{k}}^{\dagger }=(c_{\mathbf{k}\uparrow
}^{\dagger },c_{\mathbf{k+Q}\downarrow }^{\dagger })$ and passing to the
magnon representation for spin operators, we calculate the matrix electron
Green's function $\widehat{G}(\mathbf{k},E)$ to second order in the
electron-magnon interaction (this approximation corresponds to first order
in the quasiclassical small parameter $1/2S$, see \cite{IEnt}).

We do not write down the whole cumbersome expression for the matrix $%
\widehat{G}(\mathbf{k},E),$ but present the correction to the density of
states 
\begin{equation}
\delta N(E)=-\sum_{j\mathbf{k}}[\frac 1\pi \mathrm{Im}\Sigma _j(\mathbf{k,}E%
\mathbf{)/}(E-E_{\mathbf{k}j})^2+\mathrm{Re}\Sigma _j(\mathbf{k,}E\mathbf{)}%
\delta ^{\prime }(E-E_{\mathbf{k}j})]  \label{NN(E)}
\end{equation}
The self-energies are given by 
\begin{equation}
\Sigma _i(\mathbf{k},E)=\frac 12I^2\overline{S}\sum_{\mathbf{q}%
}\sum_{j,l=1,2}\{L_{\mathbf{kq}}[(-1)^{i+j+1}]+(-1)^{i+l}M_{\mathbf{kq}%
}[(-1)^{i+j+1}]\}\frac{f((-1)^lE_{\mathbf{k+q}j})+N_{\mathbf{q}}}{E-E_{%
\mathbf{k+q}j}+(-1)^l\omega _{\mathbf{q}}}  \label{sigma}
\end{equation}
where 
\begin{eqnarray}
L_{\mathbf{kq}}(\pm ) &=&(u_{\mathbf{q}}^2+v_{\mathbf{q}}^2)(1\pm I^2%
\overline{S}^2/E_{\mathbf{k}}E_{\mathbf{k}+\mathbf{q}})\pm 2u_{\mathbf{q}}v_{%
\mathbf{q}}\tau _{\mathbf{k}}\tau _{\mathbf{k}+\mathbf{q}}/E_{\mathbf{k}}E_{%
\mathbf{k}+\mathbf{q}}  \label{L} \\
M_{\mathbf{kq}}(\pm ) &=&I\overline{S}(1/E_{\mathbf{k}}\pm 1/E_{\mathbf{k}+%
\mathbf{q}})  \label{M}
\end{eqnarray}
The intra- and intersubband contributions correspond to $i=j$ and $i\neq j$.

The calculations in the narrow-band limit can be performed by using the
many-electron Hubbard operator \cite{IFTT,IKCM1} or slave boson
representations \cite{Kane}. The contribution of intersubband processes
turns out to have a different structure and does not lead to occurrence of
the singular factors $(u_{\mathbf{q}}-v_{\mathbf{q}})^2.$ In particular, the
factors of $(u_{\mathbf{q}}t_{\mathbf{k-q}}-v_{\mathbf{q}}t_{\mathbf{k}})^2$
occur which tend to zero both at $q\rightarrow 0$ and $\mathbf{q}\rightarrow 
\mathbf{Q}$ (cf. Refs.\cite{IKCM1,Kane}). The problem of interpolation
between perturbation regime and narrow-band case, which is important for
HTSC, needs further investigations.

For example, averaging $\Sigma _i(\mathbf{k},E)$ over the Fermi surface $E_{%
\mathbf{k}i}=0,$ we obtain instead of (\ref{sintr}) for the intrasubband
contribution at $T\ll |E|:$ 
\begin{equation}
\mathrm{Im}\Sigma _i^{(a)}(E)=-\frac{I^2\overline{S}}{6\pi c^3}%
F_i^{(a)}(E)\,\times \left\{ 
\begin{tabular}{ll}
$|E|(E^2+\pi ^2T^2)$ & $D=3$ \\ 
$E^2c$ & $D=2$%
\end{tabular}
\right. 
\end{equation}
with 
\begin{equation}
F_i^{(a)}(E)=\langle [L_{\mathbf{kq}}(-)-(-1)^i\mathrm{sgn}E\,M_{\mathbf{kq}%
}(-)]/\omega _{\mathbf{q}}\rangle _{E_{\mathbf{k}i}=E_{\mathbf{k+q}%
i}=0}[\omega _{\mathbf{q}}\langle \delta (E_{\mathbf{k+q}i})\rangle _{E_{%
\mathbf{k}i}=0}]_{\mathbf{q}=0}.
\end{equation}

\section{A simple scaling consideration}

The singularities of matrix elements at $\mathbf{q}\simeq \mathbf{Q}$ could
lead to the formation of marginal Fermi liquid if they would not cut at $%
E\simeq \omega ^{*}$; when taking into account the cutoff they are formally
safe. Let us consider the summation of the divergences from the dangerous
region $\mathbf{q}\simeq \mathbf{Q}$ using the ``poor-man scaling'' approach 
\cite{And1}.

We neglect here the Kondo renormalizations of the effective coupling
constant and magnon frequency, which are considered in Ref.\cite{kondo},
since they do not contain the factors of the type $(u_{\mathbf{q}}-v_{%
\mathbf{q}})^2$ and can be treated separately.

During the scaling process, the cutoff frequency is renormalized itself
owing to the renormalization of the electron spectrum, $t_{\mathbf{k}%
}\rightarrow Zt_{\mathbf{k}},$ so that $q_0\rightarrow q_0/Z,\omega
^{*}\rightarrow \omega ^{*}/Z.$

In the simplest scaling theory we have to pass to $E=0$ (effective mass at
the Fermi surface exactly). Supposing that there is no cutoff of the
dangerous electron-magnon scattering processes, we can consider the
effective mass as a function of $C$ which is the flow cutoff parameter (see,
e.g., Ref. \cite{kondo}). Usually one has to treat the limit $C\rightarrow
0. $ Here we have to remember that $\omega _{\max }>>C>>\omega ^{*}$ and to
stop scaling at the boundary of this region $\omega ^{*}$.

The electron damping owing to $s-d(f)$ interaction is determined by the
imaginary part of the polarization operator which is obtained as the
convolution of the one-electron Green's functions, 
\begin{equation}
\Pi _{\mathbf{q}}(\omega )=Z^2\sum_{\mathbf{k}}\frac{f(Zt_{\mathbf{k}%
})-f(Zt_{\mathbf{k+q}})}{Zt_{\mathbf{k+q}}-Zt_{\mathbf{k}}-\omega }
\end{equation}
Thus the quantity $\lambda _{\mathbf{Q}}$ in (\ref{IMS}), which is
proportional to the spin-wave damping, turns out to be unrenormalized.

The correction to effective electron mass in the $2D$ case according to (\ref
{z}) reads 
\begin{equation}
\delta Z^{-1}(C)=\frac 2{\pi ^2\rho c^2}\lambda _{\mathbf{Q}}\ln \left( 
\overline{\omega }/C\right)
\end{equation}
Equation for the renormalization factor $Z=Z(C\rightarrow 0)$ has the form 
\begin{equation}
1/Z=1+\frac 2{\pi ^2\rho c^2}\lambda _{\mathbf{Q}}\ln \left( Z\overline{%
\omega }/\omega ^{*}\right)
\end{equation}
This possesses the only solution with $0<Z<1$ which can be estimated as $%
Z\sim \ln ^{-1}\left( \overline{\omega }/\omega ^{*}\right) .$ Thus, despite
an appreciable renormalization of the effective mass, formation of the true
MFL state does not take place because of the presence of the cutoff.

Note that the effective mass enhancement for $3D$ ferromagnets \cite
{IKZ,IKCM2} can be treated in a similar way.

\end{document}